\documentclass[conference]{IEEEtran}
\usepackage{rotating} %

\usepackage[utf8]{inputenc}
\usepackage{graphicx}
\usepackage{graphics}
\usepackage{amssymb,amsmath,latexsym,amsfonts,amsthm}
\usepackage{booktabs}
\usepackage{array}
\usepackage{soul}
\usepackage{comment}
\usepackage[hyphens]{url}
\graphicspath{{figs/}}
\usepackage[usenames,dvipsnames]{xcolor}
\usepackage{siunitx}
\usepackage{multirow}
\usepackage{subcaption} %
\usepackage{color,soul}
\usepackage{lipsum}
\usepackage{mathtools, cuted}
\usepackage[export]{adjustbox}
\usepackage{dblfloatfix}    %
\usepackage{mathbbol}
\usepackage{bbm}
\usepackage{blindtext}
\usepackage{bm}
\usepackage{cite} %
\usepackage[english]{babel}
\usepackage{cleveref}
\usepackage{textgreek} %
\usepackage{mwe}
\usepackage{enumitem}

\usepackage{savesym} %
\usepackage{siunitx}
\savesymbol{degree}
\usepackage{gensymb} 
\restoresymbol{gensymb}{degree}

\usepackage{tabu}

\newtheorem{remark}{Remark}%
\newtheorem{proposition}{Proposition}%
\DeclareCaptionFont{mysize}{\fontsize{9pt}{12pt}\selectfont} %
\title{The Impact of Three-phase Impedances on the Stability of DER systems}

\makeatletter
\newcommand{\linebreakand}{%
  \end{@IEEEauthorhalign}
  \hfill\mbox{}\par
  \mbox{}\hfill\begin{@IEEEauthorhalign}
}
\makeatother

\author{
\IEEEauthorblockN{
Jaimie Swartz,
Alexandra von Meier\\
}

\IEEEauthorblockA{ Department of Electrical Engineering and Computer Science, \\ University of California, Berkeley, California,USA 
}

\IEEEauthorblockA{
Email: jaimie.swartz@berkeley.edu
}
}

\begin{document}
\maketitle
\begin{abstract}
In this work we explore impedance-based interactions that arise when inverter-connected distributed energy resources (DERs) inject real and reactive power to regulate voltage and power flows on three-phase unbalanced distribution grids. We consider two inverter control frameworks that compute power setpoints: a mix of volt/var and volt/watt control, and phasor-based control. On a two-bus network we isolate how line length, R/X ratio, and mutual impedances each affects the stability of the DER system. We validate our analysis through simulation of the two-bus network, and validate that the effects found extend to the IEEE 123-node feeder. Furthermore, we find that the impedance properties make it more challenging to design a stable DER system for certain placements of DER on this feeder.

\end{abstract}

\section{Introduction}
The last decade has seen rapid deployment of Distributed Energy Resources (DERs), though with simplistic controls compared to what is possible with the latest technology and policies. For example, the recent FERC Order 2222 requires regional operators to allow aggregated DERs to participate in transmission-level markets \cite{FERC_order2222}. The actions of simplistically controlled DERs and those that provide transmission-level services can introduce great variability in power flows and voltages on distribution grids. If not addressed, this can challenge power quality and reliability, for example by causing voltage flicker, increasing voltage regulator taps, desensitizing relays, or even overloading substation transformers \cite{integration_handbook}.

One standard solution to addressing power quality issues with DERs is through volt-var control. In addition, the IEEE 1547 standard \cite{IEEE_1547} now requires DERs in high-penetration areas to have volt-watt capabilities. The concern is that simulations of local droop volt-var control inverters have indicated risks of oscillations \cite{sudipta_hunting}. Therefore it is important to assess whether and under what conditions DERs that operate under volt-var with volt-watt control may cause oscillations, and whether alternatives to droop control might better regulate voltages. 

One growing class of alternatives to droop volt-var is incremental volt-var \cite{Farivar_incremental,Eggli}. The concept involves accumulating the inverter power setpoint to either minimize a local objective function \cite{Farivar_incremental}, track a voltage reference \cite{Garcia_2stage, Guannan, Helou}, or track a voltage phasor \cite{PBC_journal}. These approaches remove steady state error, thereby keeping voltages within the 5\% ANSI range. Furthermore they drive the system to an optimal state, which allows for optimal (economic) operation of the DERs while satisfying distribution-level constraints. Phasor-Based Control (PBC), which is based on the tracking of voltage phasors \cite{PBC_journal}, has been implemented with incremental volt-var control \cite{heatmap_paper}. In this work we will analyze DER systems operating under droop volt-var with volt-watt control, and PBC.

The state space model of PBC indicates that eigenvalue stability, which results in convergence as proven in \cite{heatmap_paper}, is dependent on the controller gains, the DER locations, and the network line impedances. In \cite{heatmap_paper} we explored the impact of DER locations on the DER system's stability. In this work we explore the impact of impedance properties. Three distinct metrics can characterize the impedance properties of a given network: line length, $R/X$ ratio, and mutual impedances. Through our analysis we make several remarks about the individual impact of these metrics, allowing one to predict the stability characteristics of a DER system prior to setting up a detailed simulation. 

Some works have examined the stability of volt-var and volt-watt control as impedance information is varied. However, \cite{VW_stability_2} simulates a two-bus system when varying line resistance, which changes both line length and $R/X$ ratio, rather than isolating the effect of each. Moreover, coupling between phases due to mutual impedances has been relatively unexplored in the literature apart from \cite{Garcia_2stage}, which motivates us to include 3-phase unbalanced grids. %

It is noted that impedances of real grids may be outdated or inaccurate. However, there is rich literature on estimating impedance models using sensor measurements, including distribution PMU measurements \cite{Keith_zest_journal}. Some works sidestep computing impedances explicitly and instead use power-voltage sensitivity information to design controllers \cite{sens_2}. Even though these works do not explicitly compute the impedances, because the designs implicitly account for them, it is important to investigate their role in the stability of DER systems.
\section{Problem Formulation}
\subsection{State Space Models}
We first define state space models for DER systems operating under droop volt-var with volt-watt control, and PBC.

\subsubsection{Power Flow Linearization}
\label{vmag_sec}
Consider the Distflow \cite{baran1989} branch equation for a single-phase radial network  
\begin{equation}
\left|V_i\right|^2 - \left|V_j\right|^2 = 2 (r_{ij}P_{ij} + x_{ij}Q_{ij}) + (r_{ij}^2 + x_{ij}^2) \frac{(P_{ij}^2 + Q_{ij}^2)}{\left|V_i\right|^2}, \label{DistExactMag}
\end{equation}
which approximates the relationship between voltage magnitudes $V_i, V_j$ and power flow $P_{ij}+\mathbf{j}Q_{ij}$ from node $i$ to node $j$ with complex impedance $r_{ij} + \mathbf{j}x_{ij}$.

We linearize \eqref{DistExactMag} about a nominal voltage of $1 p.u.$ by dropping the square term. Next, let $v_i$ be the squared voltage magnitude, $p_i$ the net real power, $q_i$ the net reactive power at node $i$, and define vectors $v=[v_1,v_2,...v_{n}]^T$, $p=[p_1,p_2,...p_{n}]^T$,
$q=[q_1,q_2,...q_{n}]^T$,$v_0=[v_0,v_0,...v_0]^T$ on a network with $n$ nodes. Here $v_0$ refers to the substation node which is constant at $1 p.u.$. 
 As done in \cite{Eggli,Helou}, we define the time step $k$ to be sufficiently large for the dynamics from inverters, lines, and loads to settle to steady state before new power injections are updated. 
The algebraic relationship between nodal power injections and squared nodal voltages at all nodes for time steps $k$ and $k+1$ becomes
\begin{subequations}
\begin{eqnarray}
    v_k=Rp_k+Xq_k+v_0 \label{vHelou_k}\\
    v_{k+1}=Rp_{k+1}+Xq_{k+1}+v_0 \label{vHelou_k+1}
\end{eqnarray}
\end{subequations}

where the entries of matrices R and X at the $i^{th}$ row and $j^{th}$ column are given by
\begin{subequations}\label{RXeqn2}
\begin{eqnarray}
    R_{ij}=2 \sum_{(h,k)\in \mathcal{P}_i \cap \mathcal{P}_j}^{} r_{hk}\\
    X_{ij}=2 \sum_{(h,k)\in \mathcal{P}_i \cap \mathcal{P}_j}^{} x_{hk}.
\end{eqnarray}
\end{subequations}

 $\mathcal{P}_i$ is the unique set of lines (or path) connecting node $i$ back to the substation node. 
To extend \eqref{vHelou_k} to a three-phase system we consider each phase as a separate node and triple the set of $n$ nodes as done in \cite[Appendix]{Eggli}. Each vector element in \eqref{vHelou_k} is replaced with a 3x1 vector, and each element of matrices R and X is replaced with a 3x3 block matrix. This gives $v,p,q \in \mathcal{R}^{3n \times 1}$ and $R,X \in \mathcal{R}^{3n \times 3n}$. The update equation for net nodal powers at time step $k$ is
\begin{subequations}
\begin{eqnarray}
    q_k=q^{inv}_k+q^{other}_k \label{q_update}\\
    p_k=p^{inv}_k+p^{other}_k \label{p_update}
\end{eqnarray}
\end{subequations}

\subsubsection{Droop Volt-Var with Volt-Watt State Space Model}
The droop controlled inverter update is
\begin{subequations}
\begin{eqnarray}
    q^{inv}_{k+1}=-F_{11}(v_k-\textbf{1})\\
    p^{inv}_{k+1}=-F_{21}(v_k-\textbf{1})%
\end{eqnarray}\label{dvvc_law}
\end{subequations}
where $F_{11}, F_{12}$ contain the droop curve slope parameters, and \textbf{1} is the nominal voltage  of 1pu at all nodes. We have omitted the deadband and saturation portions of the standard curve, and instead analyze DER systems that operate on the sloped portion.
Substituting the the power update equations \eqref{q_update} and \eqref{p_update} into equation \eqref{vHelou_k+1}, we have
\begin{align}
    v_{k+1}=Rp^{inv}_{k+1}+Xq^{inv}_{k+1}+R^{other}p_{k+1}+Xq^{other}_{k+1}+v_0 \label{dvvc_vupdate}
\end{align}
We subtract $v_0$ from both sides, noting that for these control laws $v_0=\textbf{1}$, then substitute control laws \eqref{dvvc_law} into \eqref{dvvc_vupdate}, giving
\begin{equation}
    v_{k+1}-\textbf{1}=(-RF_{21}-XF_{11})(v_k-\textbf{1})+Rp^{other}_{k+1}+Xq^{other}_{k+1}
\end{equation}

Let our states be the voltage magnitude tracking error $e^v=v-\textbf{1}$. We refer to the \emph{droop} closed-loop system model as
\begin{equation} \label{droop_closed_loop}
    e^v_{k+1}=\left( 0^{3n\times3n}-\begin{bmatrix} X & R \end{bmatrix}
    \begin{bmatrix} F_{11} \\ F_{21} \end{bmatrix}\right) e^v_{k} + d_{k+1} 
\end{equation}

where $d_{k+1}=Rp^{other}_{k+1} +Xq^{other}_{k+1}$ are changes in voltage magnitude from uncontrollable sources.

\subsubsection{PBC State Space Model}

The PBC model has states defined as the voltage magnitude tracking errors $e^v=v-v^{ref}$ and voltage phase angle tracking errors $e^\delta=\delta-\delta^{ref}$ at all nodes. The voltage magnitude and phase angle references $v^{ref} \in \mathcal{R}^{3n}$ tend to be near $1p.u.$ and $[0 ~-120 ~120]$ respectively. The inputs are the change in inverter power setpoints $u^q_k=q^{inv}_{k+1}-q^{inv}_k$, $u^p_k=p^{inv}_{k+1}-p^{inv}_k$ at all nodes (see our previous work \cite{heatmap_paper} for the moderl derivation). The PBC closed-loop system model is
\begin{multline} \label{PBC_closed_loop}
 \begin{bmatrix} e^v_{k+1} \\ e^\delta_{k+1} \end{bmatrix}
 =\left(  \begin{bmatrix} %
   I & 0 \\
    0 & I\\
   \end{bmatrix}
    - \begin{bmatrix} %
   X & R \\
    -\frac{1}{2}R & \frac{1}{2}X\\
   \end{bmatrix}
   \begin{bmatrix}
    F_{11} & F_{12} \\
    F_{21} & F_{22} \\
   \end{bmatrix}
   \right)
   \begin{bmatrix} e^v_{k} \\ e^\delta_{k} \end{bmatrix} \\+
     \begin{bmatrix}   c_k^q \\ c_k^p  \end{bmatrix}   +
     \begin{bmatrix}   d_k^q \\ d_k^p  \end{bmatrix}
\end{multline}

where $c_k^v=(v_{k}^{ref}-v_{k+1}^{ref})$, and $d^q_k=R(p^{other}_{k+1}-p^{other}_k) +X(q^{other}_{k+1}-q^{other}_k)$, which are the changes in voltage magnitude references and changes in voltage magnitude from uncontrollable sources, respectively. Likewise, $c_k^\delta=(\delta_{k}^{ref}-\delta_{k+1}^{ref})$, and $d^p_k=\frac{1}{2}X(p^{other}_{k+1}-p^{other}_k) -\frac{1}{2}R(q^{other}_{k+1}-q^{other}_k)$. 

For both models, the algebraic DistFlow equations and the proposed control laws form a closed-loop quasi-steady state dynamical system. The closed-loop dynamics matrix $(A-BF)$ depends on the controller gain matrix $F$ and the impedance matrix $B$. 
In previous work \cite{heatmap_paper} we designed $F$ such that the eigenvalues of $(A-BF)$ are stable, because a stable PBC system results in asymptotic phasor tracking and optimal DER operation. Here, we aim to understand how the impedance properties of $B$ affect the eigenvalues of $(A-BF)$.

\subsection{Impedance Metrics}
The impedances in $B$ can be characterized by three metrics applied to the complex impedance block $Z \in \mathcal{R}^{3 \times 3}$ for each three-phase line. 
The first is $R/X$ ratio, which is defined as
\begin{align} \label{rx_ratios}
    d_1\coloneqq \frac{R_{11}}{X_{11}}, \quad
    d_2\coloneqq \frac{R_{22}}{X_{22}}, \quad
    d_3\coloneqq \frac{R_{33}}{X_{33}}
\end{align}
The $R/X$ ratio is important for controlling DER systems because high $R/X$ ratio causes cross interactions between real power and reactive power and voltage magnitude and phase angle \cite{Sascha_upmu_app}.

The next metric is phase ratios, which we define as 
\begin{subequations}  \label{phase_ratios}
\begin{eqnarray}
    c_{x,1}\coloneqq \frac{X_{12}+X_{13}}{X_{11}}, \quad\quad
    c_{r,1}\coloneqq \frac{R_{12}+R_{13}}{R_{11}}\\
    c_{x,2}\coloneqq \frac{X_{23}+X_{12}}{X_{22}},\quad\quad
    c_{r,2}\coloneqq \frac{R_{23}+R_{12}}{R_{22}}\\
    c_{x,3}\coloneqq \frac{X_{13}+X_{23}}{X_{33}}, \quad\quad
    c_{r,3}\coloneqq \frac{R_{13}+R_{23}}{R_{33}}
\end{eqnarray}
\end{subequations}

The phase ratio is important for controlling DER systems because high mutual impedances ($R_{ij}+\textbf{j}X_{ij}~\text{for}~i\neq j$) causes cross interaction between phases. For example, high mutual impedances cause power injections on phase A to also affect voltages on phase B and C.

The last metric is line length. The magnitude of $Z$
\begin{equation} \label{linelen_singlePh}
    L_1 \coloneqq |Z|=\sqrt{R_{ii}^2+X_{ii}^2}
\end{equation}
commonly represents the length of each balanced physical conductor line in a single-line system 

A line length metric is less obvious for three-phase unbalanced systems where $Z$ is a 3x3 matrix instead of a scalar. With $\sigma_1$ as the largest singular value of $Z$, we will use
\begin{align} \label{linelen_3ph}
    L_1 \coloneqq ||Z||_2=\sigma_1(Z).
\end{align}

\subsection{Two-Bus Network}
\begin{figure}[!h]  
       \centering 
       \includegraphics[width=.28\textwidth]{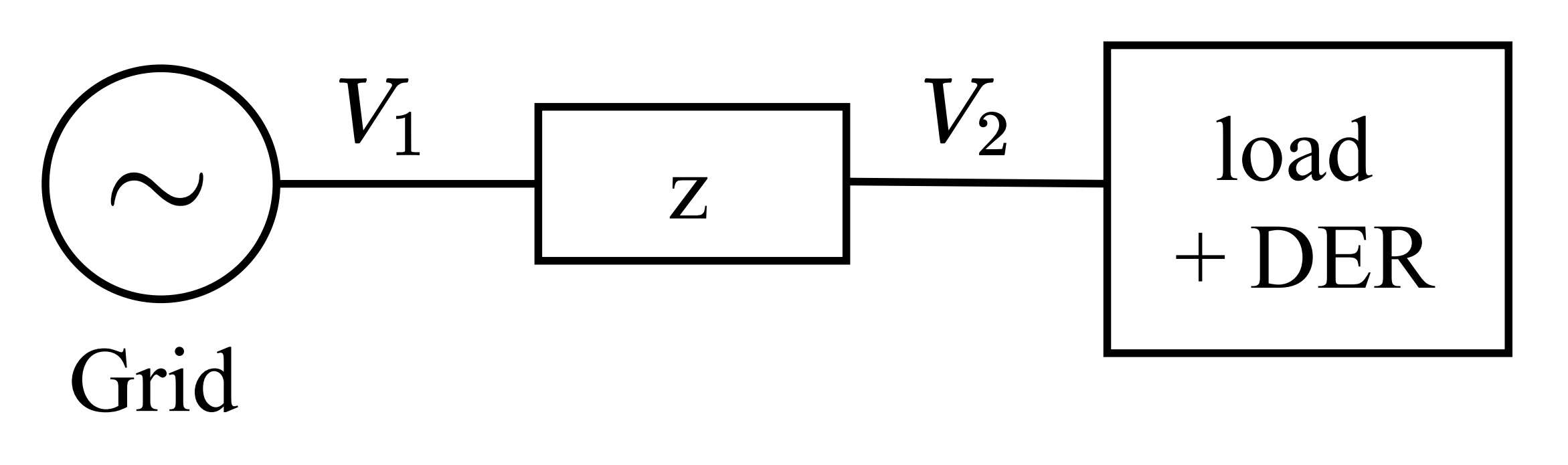}
        \caption{Slack bus connected to PQ bus. The load is constant power, and the DER power is set by PBC or droop control.}
        \label{2bus} %
    \end{figure}

For the following three scenarios we use the two-bus circuit in Fig. \ref{2bus}. We analyze how varying each impedance metric changes the dominant (maximum magnitude) eigenvalue of the closed-loop system $(A-BF)$. We define the range of stabilizing controller gains as $[0 ~...~ a_{crit}]$, where $a_{crit}>0$ is the controller gain $a$ that puts the system's dominant eigenvalue on the verge of instability. $[0 ~...~ a_{crit}]$ indicates the ease of designing a stabilizing controller, and the dominant eigenvalue predominantly determines the system trajectory. 

\subsubsection{Line Length vs. Stabilizing Controller Gains}

Consider the two-bus PBC example in Fig. \ref{2bus} as a single phase system with scalar line impedance $Z=0+\textbf{j}X$, which could represent the effective impedance of a complex circuit. Note that the line length $L_1$ is equal to the reactance $X$. For PBC, $A=I^{2 \times 2}$ and 
\begin{subequations}
\begin{eqnarray}
        B=
    \begin{bmatrix}
        X & 0 \\
        0 & \frac{1}{2}X
    \end{bmatrix}, ~F=
    \begin{bmatrix}
        a & 0 \\
        0 & 2a
    \end{bmatrix}\\
A-BF=    \begin{bmatrix}
        1-aX & 0 \\
        0 & 1-aX
    \end{bmatrix}.
\end{eqnarray}
\end{subequations}
For the droop control model, we have 
 $A=0^{2 \times 2}$, $B=[x~ 0]$, $F=[a~ a]^\top$. Thus $(A-BF)=-ax$. 
\begin{proposition} \label{linelen_prop}
    For PBC and droop control models applied to the Fig. \ref{2bus} system, increasing line length $L_1$ reduces $[0~...~a_{crit}]$ 
\end{proposition}
\noindent Proof: For the PBC case, the eigenvalues of $(A-BF)$ are $\lambda_i=1-aX~\text{for}~i=1,2$. By setting $\max_i \{ |\lambda_i| \}=1$, we solve for $a_{crit}=\frac{2}{X}$. As $X \rightarrow \infty$, $a_{crit} \rightarrow 0$. For the droop control case, $\lambda_i=-ax~\text{for}~i=1,2$, resulting in $a_{crit}=\frac{1}{X}$. As $X \rightarrow \infty$, $a_{crit} \rightarrow 0$.

\subsubsection{R/X Ratio vs. Stabilizing Controller Gains} \label{2bus_RXrat_section}
Next we isolate the effect of $R/X$ ratio \eqref{rx_ratios} on the stability of the DER system. We set $R_{ij}=0 ~\forall~ i \neq j$, $X_{ij}=0 ~\forall~ i \neq j$, $R\coloneqq R_{11}=R_{22}=R_{33}$, and $X\coloneqq X_{11}=X_{22}=X_{33}$. This makes $d_1=d_2=d_3 \coloneqq d$.

The PBC $B$ matrix is
\begin{align}
    B=
    \begin{bmatrix}
        X  & & & R& & \\
        & X & && R & \\
        & & X & & &R\\
        -\frac{1}{2}R& & &\frac{1}{2}X & & \\
         &-\frac{1}{2}R & && \frac{1}{2}X &\\
        &  &-\frac{1}{2}R && & \frac{1}{2}X\\
    \end{bmatrix}
\end{align}
For the PBC $F$ matrix, we set $F_{12}=F_{21}=0^{3\times3}, F_{11}(i,i)=a$ and $F_{22}(i,i)=2a~\text{for}~i=1,2,3$. 

For droop control, the $B$ matrix is the upper half of the PBC $B$, and $F_{11}(i,i)=a, F_{21}(i,i)=a ~\text{for}~i=1,2,3$. 

\begin{proposition} \label{rx_rat_remark}
     For PBC and droop control models applied to the Fig. \ref{2bus} system, increasing the line $R/X$ ratio (up to $R/X=1$) reduces $[0 ~...~ a_{crit}]$. 
\end{proposition}

\noindent Proof: Substituting \eqref{rx_ratios} into \eqref{linelen_singlePh} and rearranging gives $X=\frac{L_1}{\sqrt{d^2+1}}~$, and from \eqref{rx_ratios}, $R=dX$. This allows us to express $B$ in terms of $L_1$ and $d$. For the two-bus PBC system, the $(A-BF)$ eigenvalues are
\begin{align}
    \lambda_i=1-a\cdot \frac{L_1}{\sqrt{d^2+1}} \pm a \cdot d \cdot \frac{L_1}{\sqrt{d^2+1}} \textbf{j} ~\text{for}~i=1...3
\end{align}
By setting $\max_i \{ |\lambda_i| \}=1$,
\begin{align}
    a_{crit}=\frac{2}{L_1 \sqrt{d^2+1}},
\end{align}
As $d \rightarrow \infty$, $a_{crit} \rightarrow 0$.

For the two-bus droop system, $(A-BF)$ eigenvalues are
\begin{align}
    \lambda_i=-a\cdot \frac{L_1}{\sqrt{d^2+1}} - a \cdot d \cdot \frac{L_1}{\sqrt{d^2+1}} ~\text{for}~i=1...3
\end{align} with critical gain of
\begin{align}
    a_{crit}=\frac{\sqrt{d^2+1}}{L_1 (d+1)}
\end{align}
For $0\leq d \leq1$, $d$ increases $a_{crit}$ decreases. For $d>1$ as $d \rightarrow \infty$, $a_{crit} \rightarrow \infty$, but $d>1$ is less common on real grids.

\begin{figure}[!h]
   \centering 
   \includegraphics[width=.5\textwidth]{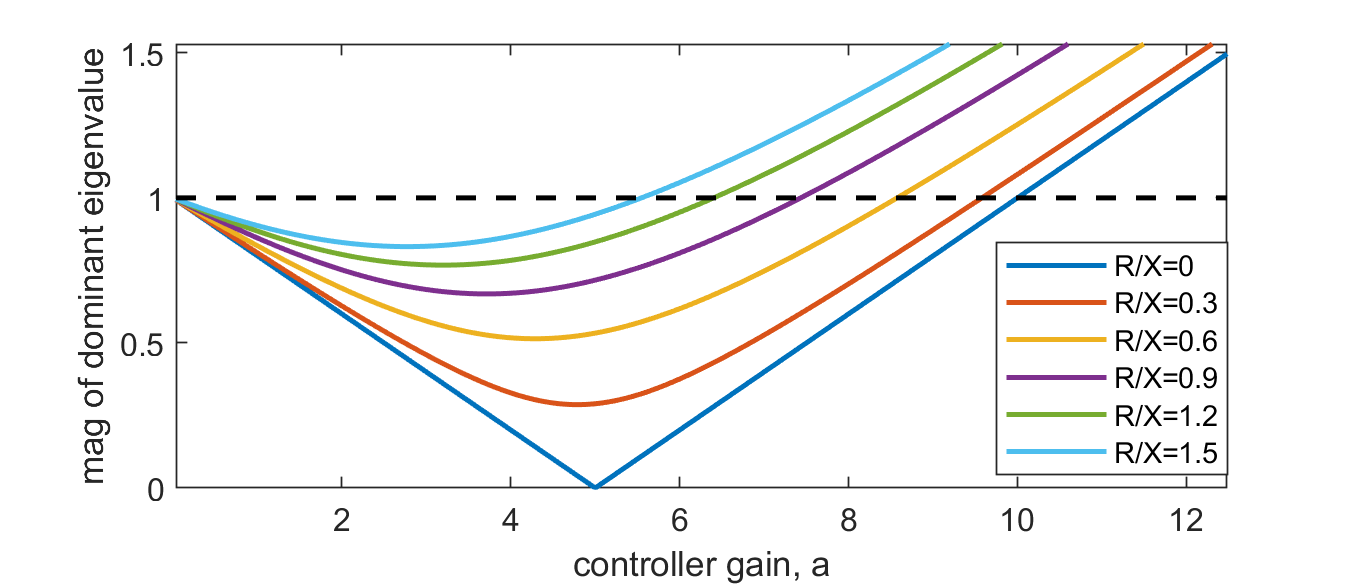}
    \caption{Magnitudes of dominant eigenvalue for two-bus PBC system as $a$ is varied, across different R/X ratios.} %
    \label{2bus_RX_matlab_pbc}
\end{figure}

\begin{figure}[!h] 
       \centering 
       \includegraphics[width=0.5\textwidth]{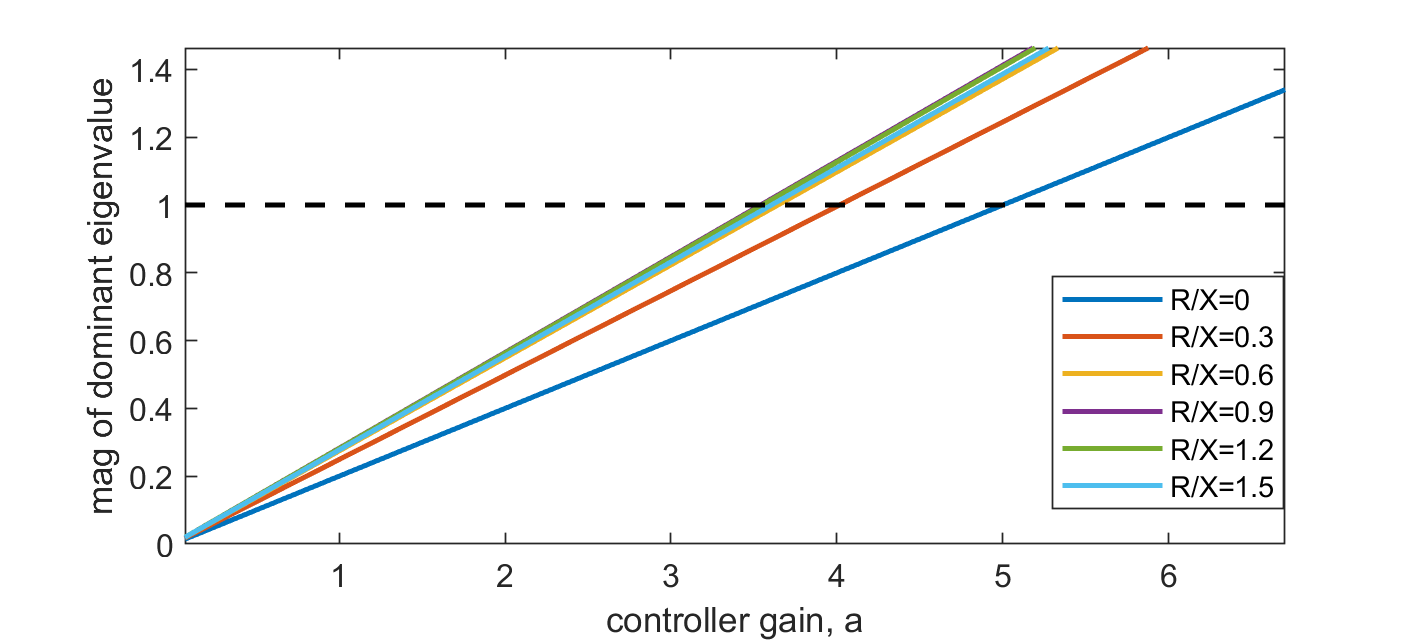}
        \caption{Magnitudes of dominant eigenvalue for two-bus droop controlled system as $a$ is varied, across different R/X ratios.} %
        \label{2bus_RX_matlab_droop}
\end{figure}

To visualize the proposition, in Fig. \ref{2bus_RX_matlab_pbc} and Fig. \ref{2bus_RX_matlab_droop} we set $L_1=0.2$, and for six R/X ratios $d$ we vary $a$ and plot the dominant eigenvalue for PBC and droop control, respectively. 

\subsubsection{Phase Ratio vs. Stabilizing Controller Gains} \label{2bus_phrat_section}
Next we isolate the effect of phase ratio \eqref{phase_ratios} on the stability of the DER system. We set $R_{ij}=0 ~\forall~ i,j$, $X\coloneqq X_{11}=X_{22}=X_{33}$, and $\bar{X}\coloneqq X_{12}=X_{13}=X_{23}$. This makes $c_{x,1}=c_{x,2}=c_{x,3} \coloneqq c_x$. The PBC $B$ matrix is
\begin{align}
    B=\begin{bmatrix}
        X  & \bar{X} & \bar{X} & & &  \\
        \bar{X} & X &\bar{X} & & &  \\
       \bar{X} & \bar{X} &X & & &\\
        &&  &\frac{1}{2}X  & \frac{1}{2}\bar{X} & \frac{1}{2}\bar{X} \\
         && & \frac{1}{2}\bar{X} & \frac{1}{2}X &\frac{1}{2}\bar{X}\\
        &  && \frac{1}{2}\bar{X} & \frac{1}{2}\bar{X} &\frac{1}{2}X\\
    \end{bmatrix}
\end{align}
and the droop control $B$ matrix is the upper half of $B$.

    \begin{figure}[!h]  
       \centering 
       \includegraphics[width=.5\textwidth]{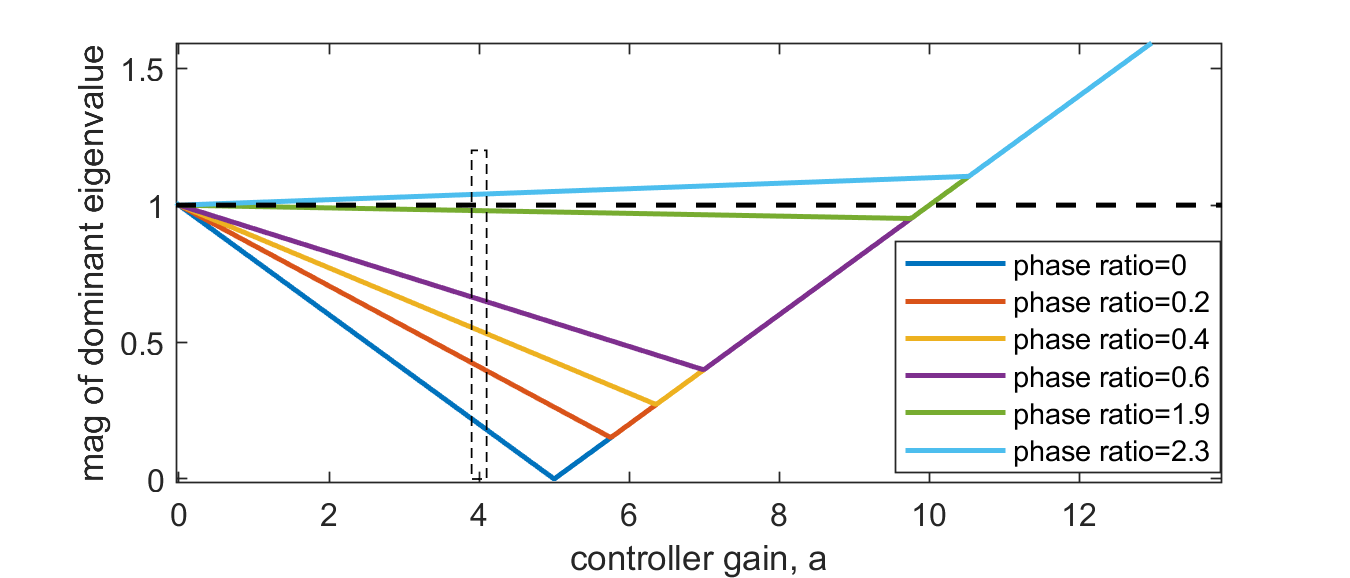}
        \caption{Magnitudes of dominant eigenvalue for two-bus PBC system as $a$ is varied, across different phase ratios.} %
        \label{2bus_phrat_MATLAB_PBC} %
    \end{figure}
    \begin{figure}[!h] 
       \centering 
       \includegraphics[width=0.5\textwidth]{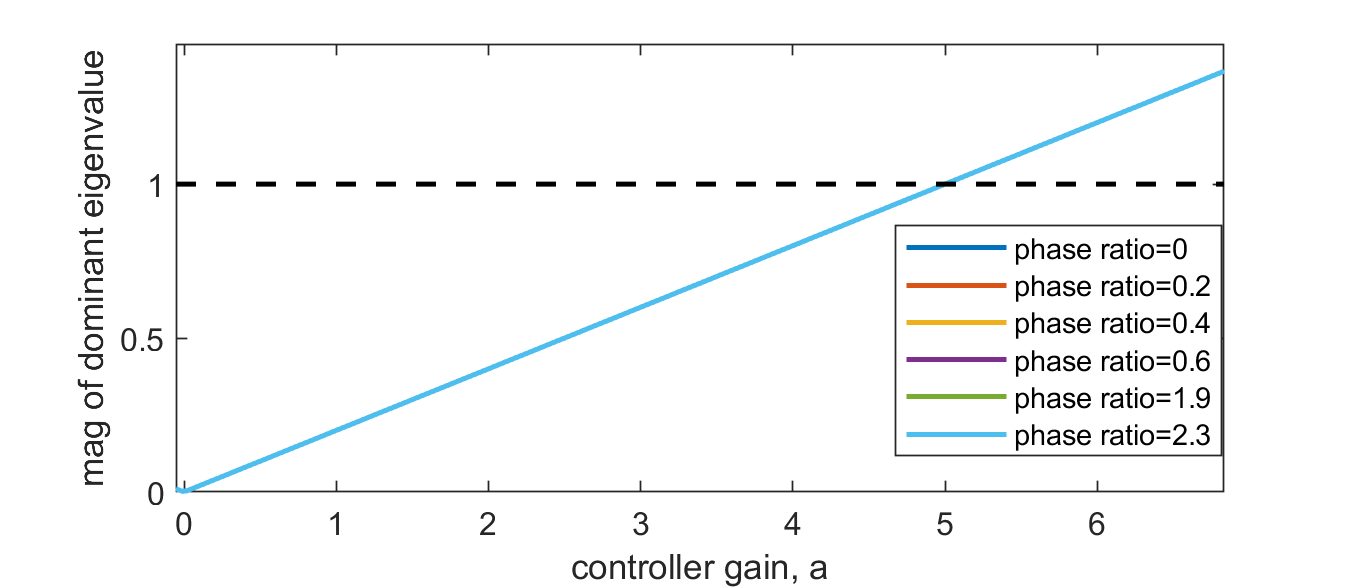}
        \caption{Magnitudes of dominant eigenvalue for two-bus droop controlled system as $a$ is varied, across different phase ratios.}  \label{2bus_phrat_MATLAB_droop}
    \end{figure}

The $F$ matrix for PBC and droop models are the same as the \ref{2bus_RXrat_section} section. For PBC, the $(A-BF)$ eigenvalues are
\begin{align}
    \lambda_i=1+a\bar{X}-aX ~\text{for}~i=1...4\\
    \lambda_i=1-2a\bar{X}-aX ~\text{for}~i=1,2,
\end{align} and for droop control they are
\begin{align}
    \lambda_i=a\bar{X}-aX ~\text{for}~i=1,2\\
    \lambda_i=-2a\bar{X}-aX ~\text{for}~i=1.
\end{align}
The three-phase line length metric \eqref{linelen_3ph} prevents us from analytically solving for $a_{crit}$ in terms of $c_x$. Instead, we choose six pairs of ($X$,$\bar{X}$) with different phase ratio $c_x$ such that $L_2=0.2$ for all pairs. We vary $a$ and plot the dominant eigenvalue $\max_i \{ |\lambda_i| \}$ for the PBC system in Fig. \ref{2bus_phrat_MATLAB_PBC} and for the droop system in Fig. \ref{2bus_phrat_MATLAB_droop}. 
We observe that 
\begin{remark}
    \begin{enumerate}[label=\alph*)]
        \item For PBC, higher phase ratios results in longer closed-loop trajectory settling time,
        \item For PBC, very high phase ratios can result in no stabilizing controller gains, \label{phrat_remark_PBC}
        \item For droop control, varying phase ratios have no effect on stabilizing controller gains or trajectory settling time \label{phrat_remark_droop}
    \end{enumerate}
\end{remark}
The first bullet and was validated by simulating the two-bus example for $a=4$ in OpenDSS, where nonlinear power flow is solved between each power setpoint update. The simulation was omitted due to space constraints. The second bullet can be seen from the phase ratio$=2.3$ case in Fig \ref{2bus_phrat_MATLAB_PBC}, and the third bullet is observed from Fig. \ref{2bus_phrat_MATLAB_droop}.

\section{Results}
\subsection{Two-Bus Simulations}
\begin{figure}[!h]  
       \centering 
       \includegraphics[width=.44\textwidth]{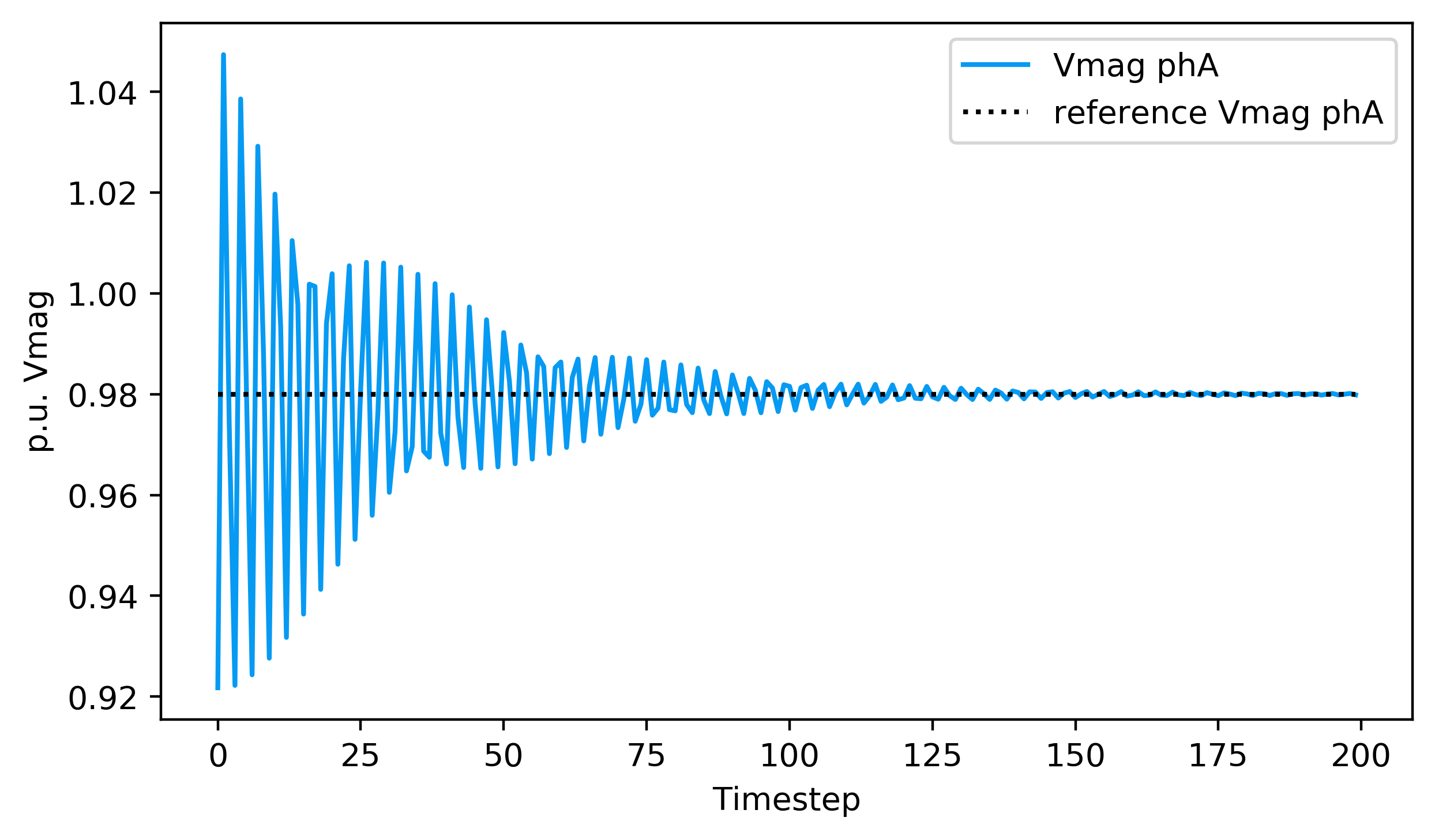}
        \caption{Instability arising immediately, for two-bus system with high R/X ratio and $a_{crit}=8.6$.} %
        \label{RX_instability} %
    \end{figure}

In this section we validate plots of Fig. \ref{2bus_RX_matlab_pbc} and Fig. \ref{2bus_phrat_MATLAB_PBC} by simulating in OpenDSS.
First we validate the PBC stabilizing control gain range for $R/X=0.6$ from Fig. \ref{2bus_RX_matlab_pbc} by simulating the single-phase two-bus network in Fig. \ref{RX_instability}. The PQ bus load is $(250 \text{kW},50 \text{kVAR})$, initial conditions are $(0.963Vp.u.,-0.0395 \degree)$, and phasor reference are $(0.98 Vp.u., -0.036 \degree)$.  The critical gain from the nonlinear simulation matches the one found by the linear analysis ($a_{crit}=8.6$). For this gain, we observe that unstable oscillations arise immediately.

    \begin{figure}[!h]  
       \centering 
       \includegraphics[width=.47\textwidth]{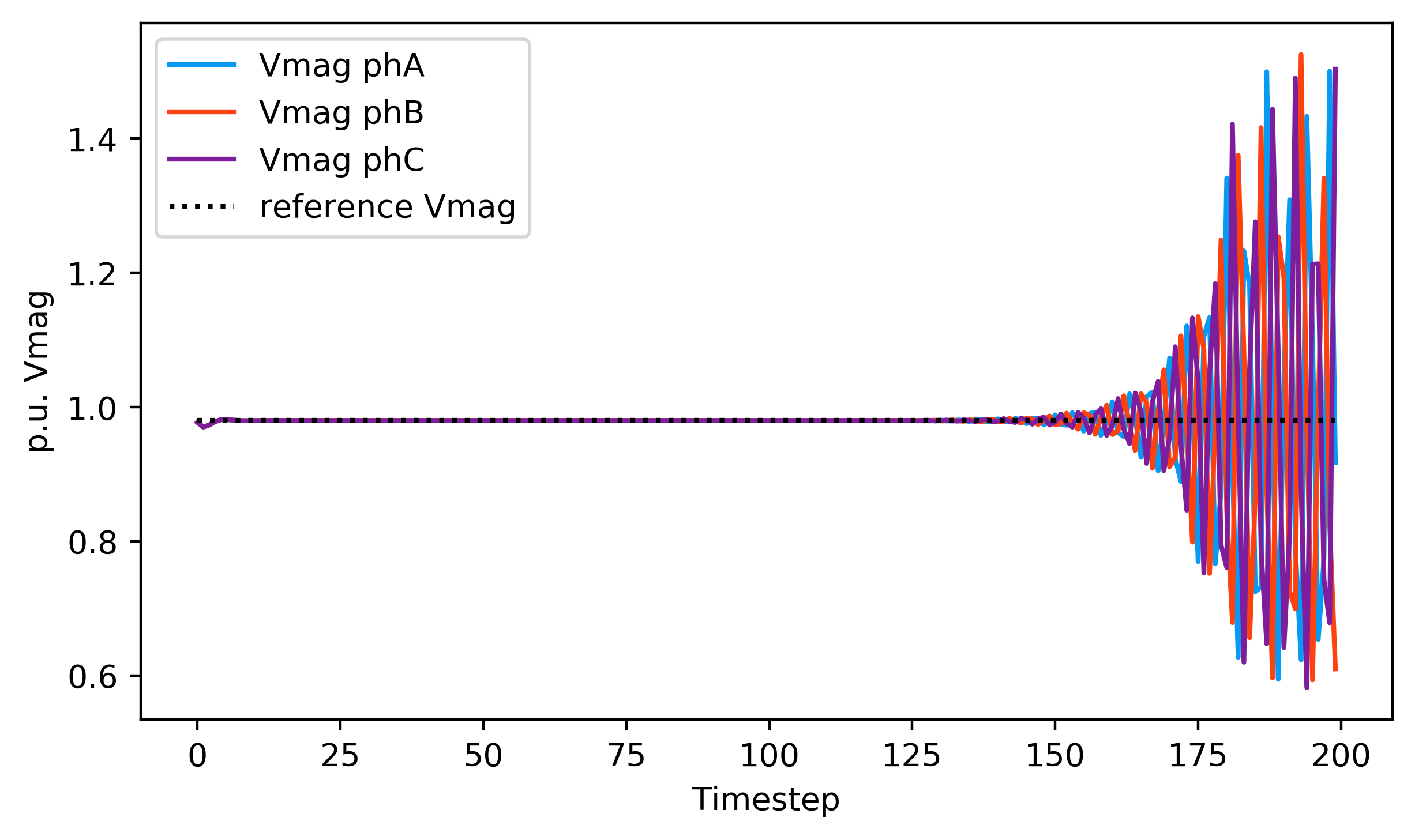}
        \caption{Instability arising after much delay, for two-bus system with high phase ratio and $a_{crit}=13$.} %
        \label{ph_instability} %
    \end{figure}

Next we validate the PBC stabilizing control gain range for $R/X=0.6$ from Fig. \ref{2bus_phrat_MATLAB_PBC} by simulating the three-phase two-bus network in Fig. \ref{ph_instability}. The PQ bus load, initial conditions, and phasor references are the same as the previous paragraph, though applied to three phases. The critical gains from the nonlinear simulation matches the one found by the linear analysis ($a_{crit}=10$). For this gain, unstable oscillations begin after 150 seconds. This delay is concerning, indicating that oscillations from high phase ratio lines may be more difficult to detect than oscillations from high R/X ratio lines.
\subsection{Experiment Setup on Larger Feeders}
\subsubsection{Disturbance Rejection Requirement for Droop Control}
While stability of the droop model \eqref{droop_closed_loop} ensures that no or in the worst case damped voltage oscillations occur, stability is not sufficient to guarantee the voltage magnitudes are driven inside the 5\% ANSI range. This is because droop control admits a steady state error, which may land outside the 5\% range depending on the system's operating point.

 One could evaluate a set of droop control parameters by how much the steady state error changes after a voltage step change disturbance $\triangle V^d \in \mathcal{R}^n$. \cite{adaptVVC} derives the change in steady state error to be $\triangle SS=[I+BF]^{-1}\triangle V^d$, with $B$ and $F$ defined in \eqref{droop_closed_loop}. We see that when $F=0^{6n \times 3n}$, any disturbance will shift the steady state error by just as much ($\triangle SS=\triangle V^d$). Nonzero control gains \emph{contracts} the steady state error, thereby providing some disturbance rejection capability.

\subsubsection{About the Heatmap Tool}

In our previous work \cite{heatmap_paper}, we incorporated the PBC model \eqref{PBC_closed_loop} into a heatmap tool that visually illustrates good locations for configurations of DERs (actuator nodes) and associated sensors (performance nodes). The tool's code can be downloaded at: \\\noindent \url{https://github.com/jaimiosyncrasy/heatMap}. In this work, we extended the tool to work for the droop model \eqref{droop_closed_loop}. 

The locations of actuator-performance node pairs (APNPs) enforces certain structural requirements on $F$ per configuration. An APNP is \emph{co-located} if a DER and its associated sensor are at the same node. Additionally, the $F$ matrix structure is simplified by realistic assumptions about the configurations: all droop 3x3 blocks in $F$ representing three-phase nodes are diagonal, and for PBC $F_{12}=F_{21}=0$. See Assumptions 1 and 2 of \cite{heatmap_paper} for details.

 For a given configuration operating under droop control, if we can find at least one $F$ such that (1) the closed-loop system is stable in the sense of Lyapunov, and (2) the steady state error contracts by at least 7\% when $[\triangle V^d]_i=0.535~Vp.u.~\forall~i=1...n$, our tool indicates that the configuration and associated $F$ is \emph{good}.
For a configuration operating under PBC to be \emph{good}, only requirement (1) is needed.

\subsection{Validation of the Two-Bus Remarks on Larger Feeder}
We now determine whether Proposition \ref{rx_rat_remark} and remark 1 found for the two-bus network extend to the unbalanced IEEE 123-node feeder (123NF) \cite{IEEE_test_feeders}. We use the heatmap tool to find an $m$-node configuration under PBC or droop control that is \emph{good} on the original 123NF. Once found we record the best associated $F$, which is the $F$ found that corresponds $(A-BF)$ having the smallest dominant eigenvalue magnitude. Then we apply this best $F$ to the same configuration on a modified 123NF, where every line has a higher R/X or phase ratio with the same line length $L_2$ \eqref{linelen_3ph}. If the configuration on the higher-ratio 123NF is no longer \emph{good}, that result supports the remarks. Then we do the reverse. That is, we find a \emph{good} configuration on the modified 123NF and apply the best associated $F$ to the original 123NF, noting whether the system is no longer \emph{good}. If no longer good, that result contradicts the remarks. 

Each entry of Table \ref{tab:1.5x_table} records the percentage of results that support, with respect to a total of ten results that support or contradict the remarks. 
The experiments involve the original 123NF versus the 123NF that is modified to have the R/X ratio or phase ratio of every line multiplied by 1.5x.
The first two columns of Table \ref{tab:1.5x_table} supports Proposition \ref{rx_rat_remark}. The third column supports remark 1.\ref{phrat_remark_PBC}. The last column, where phase ratios equally often support and contradict in the experiment, supports remark 1.\ref{phrat_remark_droop}.

\begin{table}[htbp]
\caption{\label{tab:1.5x_table} Percent of configurations that lose stability when $R/X$ or phase ratio ratio is increased. 'N/A' indicates no stable configurations were found.}
\begin{center}
\begin{tabular}{|p{0.8cm}|p{1.3cm}|p{1.3cm}|p{1.3cm}|p{1.3cm}|}
    \hline
    \multirow{2}{*}{} &
      \multicolumn{2}{c|}{R/X ratio} &
      \multicolumn{2}{c|}{phase ratio} \\ \hline
 m & PBC, orig $\leftrightarrow$ 1.5x & droop, orig $\leftrightarrow$ 1.5x &  PBC, orig $\leftrightarrow$ 1.5x & droop, orig $\leftrightarrow$ 1.5x\\ \hline
1 & 100\% & inconc. & 100 & 50\% \\ \hline
5 & 100\% & 100\% & 100\% & 60\%\\ \hline
10 & 100\% & 100\% & 100\% & 60\%\\ \hline
15 & N/A & 100 & N/A\% & 50\% \\ \hline
\end{tabular}
\end{center}
\end{table}

\subsection{Branch Comparison on 123NF}
In order to conduct an experiment that compares feeder branches with differing R/X and phase ratios, we first describe one of the heatmap tool's placement process.

\subsubsection{Co-located Placement Process (CPP)}
The CPP illustrates good places to place the next co-located APNP to maintain stability of the DER system. It iterates through every empty node in the feeder, fixing each as a candidate co-located APNP. Then, we generate a heatmap on the network, where a node's color indicates whether the configuration created by appending the candidate APNP to the existing set of APNPs is \emph{good}. If the node is blue, several (at least 7\% of) $F$ matrices sampled from the parameter space defined in \cite{heatmap_paper} make the configuration \emph{good}. If the node is yellow, only a few $(< 7\%)$ were found, and if red, no $F$ matrices were found.

\subsubsection{Branch Comparison Experiment} 

    \begin{figure}[!h]  
       \centering 
       \includegraphics[width=0.4\textwidth]{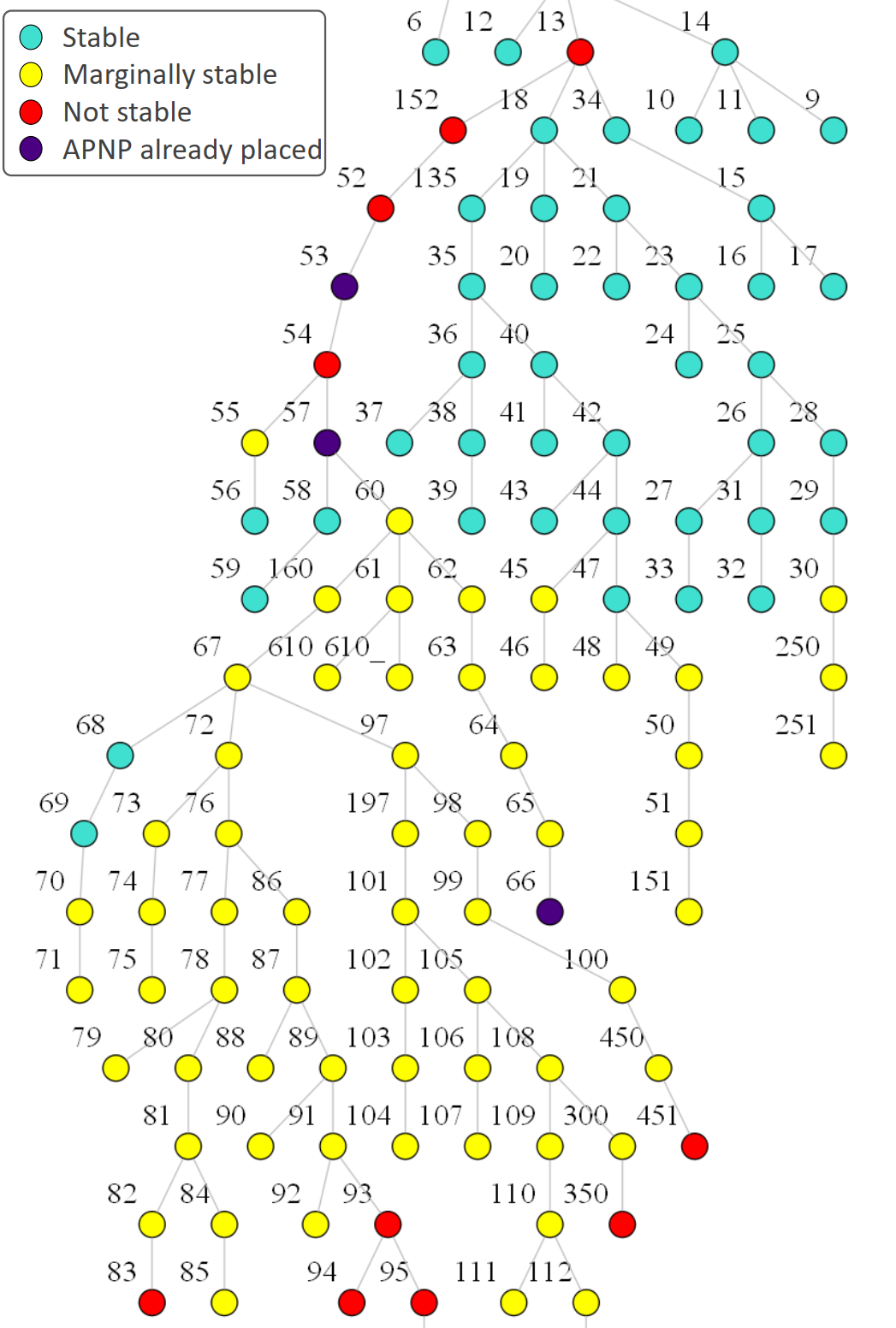}
        \caption{PBC heatmap of configuration $\chi_1$ , which contains the bad branch of node 62 to 66. The blue, yellow, and red nodes shown are tallied into the 'PBC, high ratio branch' bar graph of Fig. \ref{bar_graph}.}
        \label{CPP_heatmap} %
    \end{figure}
    
We define two configurations of co-located APNPs, $\chi_1=$
[node\_8, node\_53, node\_57, node\_66] and $\chi_2=$[node\_8, node\_53, node\_57, node\_74]. Note that $\chi_1$ has an APNP on branch [node\_62 node\_66], which has the highest three-phase R/X ratios and phase ratios, as measured by each node's path to the substation. 

\begin{figure}[!h]  
       \centering 
       \includegraphics[width=0.45\textwidth]{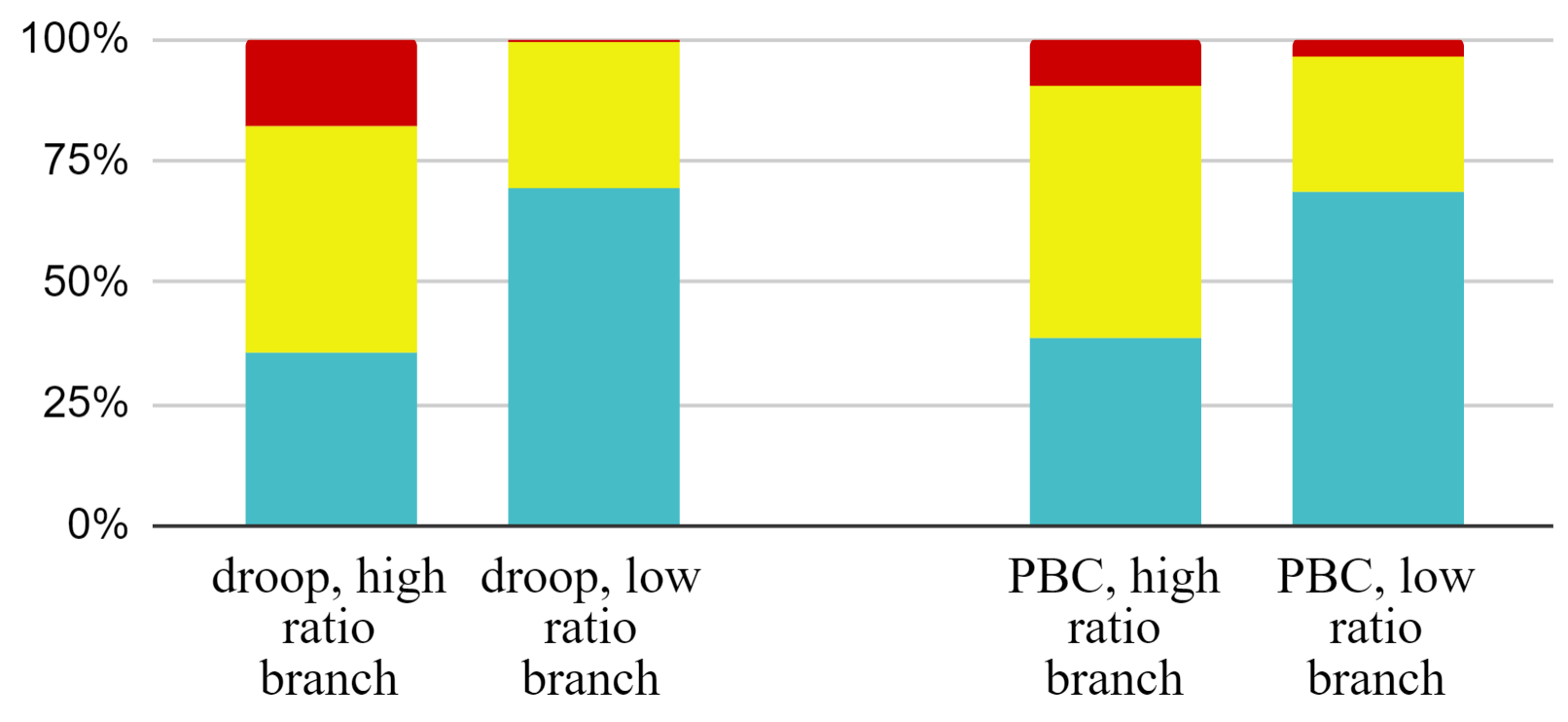}
        \caption{Heatmap color breakdown when configuration has a co-located DER on the high ratio branch (config $\chi_1$) vs. on the low ratio branch (config $\chi_2$).}
        \label{bar_graph} %
    \end{figure}

We place $\chi_1$ on the 123NF operating under PBC, then run the CPP to find good locations for the fifth APNP, which is illustrated by the tool's heatmap in Fig. \ref{CPP_heatmap}. 
We run the CPP for thee more variations: $\chi_2$ controlled with PBC, droop controlled with $\chi_1$, and droop controlled with $\chi_2$. We tabulate the number of blue, yellow, and red nodes across the four runs in Fig. \ref{bar_graph}. The greater percentage of red nodes for both droop and PBC models indicates that placing co-located DER-sensor pairs on high-ratio branches makes it difficult to place subsequent DERs.

\section{Conclusion}
\label{sec:Conclusion}
In this work we investigated the impact of network impedances on the stability of DER systems operating under droop and PBC models. We proved that long lines and high R/X ratios reduce the range of stabilizing controller gains (Proposition 1 and 2). Then, for PBC we found that high phase ratios reduce closed-loop settling time and can result in no stabilizing gains. We validated our 2-bus system analysis with simulations, and verified that the effects extend to the 123NF. Finally, we demonstrated that placing co-located DER-sensor pairs on high R/X and phase-ratio branches of the 123NF makes it difficult to place subsequent DERs.

In future work we will extend the branch comparison experiment to other large feeders. We will also accompany the analysis with large-scale simulations to validate the stability results. Finally, we seek to theoretically prove the observed relationship between high ratio branches and the ease of finding stabilizing controller gains.

\section*{Acknowledgements}
We thank Daniel Tutt and Keith Moffatt for their insightful discussions. This work was supported by the U.S. Dept. of Energy, Award DE-EE0008008. 

\bibliography{library.bib}
\bibliographystyle{ieeetr}

\end{document}